\documentstyle[aas2pp4]{article}
\received{}
\revised{}
\accepted{}
\slugcomment{(ApJ in press)}

\lefthead{Sahu et al.}
\righthead{A UHRF study of two high galactic latitude stars}

\newcommand{\vlsr}{$v_{\mbox{\tiny LSR}}$}

\newcommand{\Tk}{$T_{\mbox{\rm k}}$}
\newcommand{\etal}{et al.}
\newcommand{\kms}{$\mbox{km\,s}^{-1}$}
\newcommand{\ang}{\,\AA}

\slugcomment{(ApJ in press)}

\begin{document}
\title{ Atomic and molecular interstellar absorption lines toward  \\
 the high galactic latitude stars HD~141569 and HD~157841 at ultra-high resolution}

\author{M. S. Sahu}
\affil{Goddard Space Flight Centre, Code 681, Greenbelt, MD 20771,
msahu@pulsar.gsfc.nasa.gov}

\author{J. C. Blades, and L. He\altaffilmark{1}}
\affil{Space Telescope Science Institute, 3700 San Martin Drive, Baltimore, 
MD 21218, blades@stsci.edu, lida@charon.phys.rpi.edu}

\author{Dap Hartmann}
\affil{Center for Astrophysics MS 72, 60 Garden Street, Cambridge, MA 02138
dap@abitibi.harvard.edu}

\author{M. J. Barlow and I. A. Crawford}
\affil{Department of Physics and Astronomy, University College London, 
Gower Street, London, WC1E 6BT, UK, iac@star.ucl.ac.uk, mjb@star.ucl.ac.uk}

\altaffiltext{1}{on leave from Physics Department, Rensselaer Polytechnic 
Institute, Troy, NY 12180}


\begin{abstract}

We present ultra-high resolution (0.32 km s$^{-1}$) spectra
obtained with the 3.9m Anglo-Australian Telescope (AAT) and Ultra-High-Resolution Facility (UHRF), of interstellar Na~{\sc i} D$_1$, 
Na~{\sc i} D$_2$, Ca~{\sc ii} K, K~{\sc i} and CH absorption toward two high galactic latitude stars 
HD~141569 and HD~157841. We have compared our data with 21-cm observations obtained from the Leiden/Dwingeloo H~{\sc i} survey. 
We derive the velocity structure, column densities of the clouds represented by the various components and identify the clouds with ISM structures seen in the region at other wavelengths. We further derive abundances, linear depletions and H$_2$ fractional abundances for these clouds, wherever possible.

Both stars are located in regions of IRAS 100$\mu$m emission associated with high galactic latitude molecular clouds (HLCs) : HD~141569 lies, in projection, close to MBM~37 and the Lynds dark cloud L~134N while HD~157841 is in the vicinity of the MBM~151. 

Toward HD~141569, we detect two components in our UHRF spectra : a weak,
broad $b$ = 4.5 \kms\ component at -- 15 \kms\ , seen only in Ca~{\sc ii}
K absorption and another component at 0 \kms\ , seen in Na~{\sc i} D$_1$, Na~{\sc i} D$_2$, Ca~{\sc ii} K, K~{\sc i} and CH absorption. The cloud represented by the -- 15 \kms\ component, is warm and may be located in a region close to the star. The cloud represented by the 0 \kms\ component has a Ca
linear depletion $\delta$(Ca) = 1.4 $\times$ 10$^{-4}$ and shows evidence for the presence of dust, consistent with strong 100$\mu$m emission seen in this region. The H$_2$ fractional abundance  $f$(H$_2$), derived for this cloud is 0.4, which is typically what is observed toward HLCs. We conclude that this 0 \kms\ cloud is associated with MBM~37 and L~134N based on the presence of dust and molecular gas (CH) and good velocity agreement with CO
emission from these two clouds. This places HD~141569 beyond MBM~37 and
L~134N, which are estimated to be at $\approx$ 110 pc. 

In the case of the HD~157841 sightline, a total of 6 components are seen on our 
UHRF spectra in  Na~{\sc i} D$_1$, Na~{\sc i} D$_2$, Ca~{\sc ii} K, K~{\sc i} and CH absorption. 2 of these 6 components are seen only in a single species.
The cloud represented by the components at 1.85 \kms\ has a Ca linear depletion
$\delta$(Ca) = 2.8 $\times$ 10$^{-4}$, indicating the presence of dust. The 
$f$(H$_2$) derived for this cloud is 0.45 and there is good velocity agreement with CO emission from MBM~151. To the best of our knowledge, this 1.85 \kms\ 
component towards HD~157841 is the first one found to have relative line widths
that are consistent with pure thermal broadening only.
We associate the 1.85 \kms\ cloud seen
in our UHRF spectra with MBM~151 and conclude that HD~157841 must lie beyond
$\sim$ 200 pc, the estimated distance to MBM~151.

\end{abstract}



%

\section{Introduction}
In this paper, we report the results of an ultra-high-resolution (0.32 \kms) absorption line study of two stars, HD~141569 and HD~157841. Both stars were chosen as background sources for detailed spectroscopic studies of high galactic latitude molecular clouds (HLCs), as well as gas in the nearby halo.

HLCs show structure, with dense clumps 
as small as 0.03 pc, which appear to be virially unstable (Pound et al. 1990). 
Some of these clouds appear to be embedded in the local hot 
interstellar medium as suggested by the probable detection of X-ray shadowing
from two MBM clouds (Burrows and Mendenhall, 1991). If this is indeed the case,
it is not clear what physical process or combination of processes is causing 
such enhanced molecular abundances. This has led some authors (Blitz, 1990) to 
conclude that HLCs may represent a completely distinct population of clouds.
More recent work by Gir et al. (1994) indicates that HLCs
are almost always located along filamentary or looplike HI structures and appear to have condensed from atomic gas in situ rather than having been entrained in the HI.

These clouds are an important factor both in the composition and the energy balance 
of the nearby interstellar medium. They have molecular masses about 50 M$_{\odot}$
and the molecular gas contribution of these high latitude clouds in the solar 
neighborhood up to 100 pc, is about 5000  M$_{\odot}$ (Magnani 1993).
The relation and exchange of energy between the molecular gas in these clouds 
and the ambient highly ionized, moderately ionized and atomic gas is
unclear although it has been suggested by Blitz (1990) that compression of these
clouds could lead to a phase transition of atomic hydrogen into molecular hydrogen.

Absorption line studies toward stars whose lines of sight intercept regions close to HLCs, probe the diffuse component of the ISM associated with these high latitude clouds. Typical values of the kinetic temperatures for the diffuse component of the ISM range between 45 to 130 K (Savage et al. 1977). To resolve absorption lines 
arising from clouds with cloud kinetic temperatures of \Tk\ = 50 K, a  velocity 
resolution of 0.30 \kms\  corresponding to a resolving power R = 10$^6$ is necessary.  The Ultra High Resolution Facility 
(UHRF)  at the Anglo Australian Telescope (AAT) is currently the world's highest-resolution 
optical astronomical spectrograph (Diego et al., 1995, Barlow et al., 1995)
and is ideal for studies of such cool, interstellar clouds. 

We have begun a study of 
the spatial distribution and properties of atomic and molecular interstellar species 
towards HLCs at ultra high resolution (velocity resolution = 0.32 \kms\ ), 
and 
have compiled a list of background stars  with reliable photometric and 
spectroscopic data and lying in the line of sight to HLCs.   We have already 
observed a small fraction of this sample, utilizing the UHRF (Blades et al. 1997), and in the following sections, we present UHRF observations toward HD~141569 and HD~157841.

\subsection{Overview of the HD~141569 sightline}

HD~141569 has been variously classified as an isolated Herbig Ae/Be star (Th$\acute{e}$ et al. 1994) and a B9.5V star by Penprase (1992) and an AOVe
star by Dunkin et al. (1997). It is an emission-line star associated with an IRAS source, IRAS~15473-0346, which is not exactly coincident with it. The heliocentric radial velocity of the star is --6.4 \kms\ (Frisch, 1987). In the direction of HD~141569, which is located in galactic coordinates at ($l$, $b$) = (4.$^\circ$2, +36.$^\circ$9), this translates into a LSR radial velocity of --20.1 \kms\ . The star exhibits 
excess far-IR emission (Oudmaijer et al. 1992) believed to arise due to emission from dust grains present in a circumstellar disk, and Sylvester et al. (1996) include it in their list of Vega-like systems. 

A high galactic latitude dark cloud complex at ($l$, $b$) $\sim$ (4$^\circ$,
36$^\circ$) lies close to the star in projection. This cloud complex includes the Lynds dark clouds L~134, L~183 (which is also often referred to as L~134N; henceforth we will refer to this cloud as L~134N) and L~1780. The dense molecular CO core MBM~36 is
located within L~134, while the CO core in L~134N is referred to as MBM~37 (Magnani et al. 1985). Of these two dark clouds, L~134N, centered at 
($l$, $b$) = (6$^\circ$.0, 36$^\circ$.8) and 2$^\circ$ north of L~134, lies closest to HD~141569. HD~141569 lies well within the 100$\mu$m emission contours of the 
L~134N dark cloud (Laureijs et al. 1991). 

Penprase (1992) estimated a distance of 190$\pm$110 pc for HD~141569 based on spectroscopic and UBV and Str\"{o}mgren
photometric data. There are various distance estimates to the L134, L134N and L1780 complex: 100$\pm$50 pc, based on optical extinction and surface brightness
observations by Mattila (1979), 110$\pm$10 pc based on photometric data by Franco (1989) and 160 pc  based on reddenings of a large number of stars in this direction by Snell (1981). The parallax of HD~141569 from the Hipparcos Catalogue is 10.10$\pm$0.83~mas, corresponding to a distance of 99$\pm$8~pc.
This implies that the distance to the star is very similar to that of the L~134,
L~134N and L~1780 complex. Penprase (1992) estimated an E(B--V) of 0.12 $\pm$ 0.09 for HD~141569. The Hipparcos Catalogue lists Tycho magnitudes which transform to give (B--V) = 0.078$\pm$0.007, which for an A0V star corresponds to an E(B--V) of 0.10$\pm$0.03 (the uncertainty of 0.03 is due to the uncertainty in the (B--V)$_{o}$ calibration for dwarf stars). The star falls just outside the outermost CO contours which trace MBM~37, the CO core in L~134N (Caillault et al. 1995). Based on available information, although HD~141569 appears to lie approximately at the distance of L134N
and clearly does not lie within the dark cloud itself, it is not certain whether it lies in front of, or behind this dark cloud. This point is discussed further
in $\S$ 3.1.3. and our UHRF data indicate that HD~141569 lies beyond L~134N and MBM~37.

\subsection{Overview of the HD~157841 sightline}

HD~157841, a B9 star (SIMBAD database) at ($l$, $b$) = (16$^\circ$.8, 15$^\circ$.7), lies in projection near the HLC, MBM~151,
which is located at ($l$, $b$) = (21$^\circ$.5, 20$^\circ$.9). We have searched the SIMBAD database and related literature and have been unable to
find any radial velocity measurement for this star.\\

Based on CO velocity agreement, Penprase (1993) suggested that MBM~57
located at ($l$, $b$) = (5$^\circ$.1, 30$^\circ$.8) is related to MBM~151, which is $\sim$ 19$^\circ$ away. Photometric data of stars in the region
around MBM~57 (Franco, 1989), show an increase in extinction beyond 100 pc.
Penprase (1992) also observed this increase in the reddening and estimated a distance of 150 to 210 pc for the near edge of MBM~57. Assuming that MBM~57
and MBM~151 are related, the distance to MBM~151 is $\sim$ 200 pc. The CO (J= 1$\rightarrow$0) LSR velocity of MBM~151 is --0.8 \kms\ (Magnani et al., 1985). The region near HD~157841 has not been mapped in CO and it is not known if there is any CO emission present at the position of the star or in its vicinity.
The Hipparcos Catalogue lists a parallax for HD~157841 of 5.70$\pm$1.16~mas
corresponding to a distance of 175$\pm$$\stackrel{45}{_{30}}$~pc. It also lists 
Tycho magnitudes which transform to yield (B--V) = 0.213$\pm$0.008. For a
B9V star, this corresponds to an E(B--V) of 0.28$\pm$0.03. In $\S$ 3.2.5., we identify the various components seen in the UHRF spectra toward HD~157841.

\section{The data and its reduction}
\subsection{The UHRF data}
The spectra of  HD~141569 and HD~157841 were obtained during April 25 to 28, 1994 and July 14 and 15, 1995, using the UHRF at the Coud\'{e} focus of the 3.9-meter Anglo-Australian
Telescope. For details about the UHRF refer to Diego et al. (1995). The configuration of the spectrograph was set to give close to the
highest achievable resolution which is R = 938,000 $\pm$ 15,000. We used the 35-slice image slicer (Diego, 1993)
at the entrance aperture. We masked the image slicer down to give
18 slices on the detector which allowed for the simultaneous
detection of the inter-order background on either side of the stellar
spectrum. 

The detector was a Tektronics CCD (1024 $\times$ 1024) with a pixel size of 24 $\times$ 24 $\mu$m. The width of
each individual slice produced by the image slicer is about
30 pixels, and the 18 slices filled over half the detector format while the
order background filled the remainder. The entire CCD was read out
with eight times binning in the direction perpendicular to the
dispersion.

The instrumental set-up gave a wavelength coverage of 2.5\AA,
requiring separate exposures for each of the Na~{\sc i} D lines. Individual exposures were normally  1800 seconds, and
the total integration times were 30 minutes each for the D$_1$ and D$_2$ spectra, 60 minutes for Ca~{\sc ii} K and 180 minutes each for the K~{\sc i} and CH spectra. Comparison exposures using a Th-Ar lamp were used to establish
the wavelength scale and showed that the plate scale is linear with
wavelength over the 2.5\ang\ coverage (Barlow \etal\
1995). Measurements of the 6328.160\ang\  stabilized He-Ne laser line taken 
during
the observing run confirmed that we achieved a resolution of 0.32 \kms\
(FWHM).

Standard IRAF reduction packages were used to combine individual integrations  after cosmic ray events and the inter-order background had been removed. A scattered light correction
of 3.5\% was applied to all the data, derived from observations of a
highly saturated D line in a different star (HD~24263) recorded during the same
period. The scattered-light fraction for the Ca~{\sc ii}, CH and K~{\sc i} lines
is not the same as for the Na~{\sc i} D$_1$ and D$_2$ lines. In particular
for the Ca~{\sc ii} K line, inspite of background subtraction (by taking 
averages of the inter-order regions above and below the stellar spectrum), there may still be a residual of as much as 10\% (Crawford et al. 1994). We do not have observations of saturated Ca~{\sc ii} K lines as in the case of the Na~{\sc i} D lines
and therefore do not have exact values for the scattered light fraction in the 
Ca~{\sc ii} line. However, we have compared our total  equivalent widths with the values published by Penprase (1993), assuming that the Penprase data has no scattered light contribution. The equivalent widths obtained for Na~{\sc i}
and CH for both HD~141569 and HD~157841 are consistent with zero contribution from scattered light. For the case of the Ca~{\sc ii} line, we estimate the contribution of scattered light to the measured equivalent widths to be $\sim$ 
15\% for HD~141569 and $\sim$ 8\% for HD~157841. For the case of the K~{\sc i}
lines, we are not able to make such scattered light estimates since there are
no published K~{\sc i} equivalent widths for either star.
We have not subtracted this correction for the Ca~{\sc ii} lines and estimate that the uncertainty due to scattered light has a small effect on the 
derived column densities ($\le$ 10\%) and has almost no effect on the measured linewidths, which is our main concern in this paper.

The spectra were normalized by fits of straight lines to
sections without telluric or interstellar absorption.  Telluric
absorptions around the D line region were removed by dividing by a 
UHRF telluric template spectrum of $\alpha$~Eri, which was rescaled so that 
the  
depth of individual telluric features matched those of unblended telluric 
lines in the spectra. Rest wavelengths in air of
5895.9234\ang\ for D$_1$,  5889.9512\ang\ for D$_2$, 3933.663~\ang\ 
for Ca~{\sc ii} K, 7698.974~\ang\ for K~{\sc i} and 4300.3133~\ang\ for CH, were adopted from Morton (1991).   The signal-to-noise
ratios derived from the r.m.s. fluctuation in the stellar continuum  
for HD~141569 are 20 for D$_1$, 28 for D$_2$ and 10 for the Ca~{\sc ii} K, CH and K~{\sc i} spectra.
In the case of HD~157841, the S/N values are 35 for D$_1$, 16 for D$_2$, 10 for K~{\sc i}, 20 for Ca~{\sc ii} K and 42 for CH.

\subsection{Profile fitting}

We used the method of profile fitting developed by Welty \etal\ (1994)
to derive the properties of the individual interstellar clouds. The
essential assumptions are that each cloud is homogeneous and can be
described by a column density $N$, radial velocity $v$, and a Gaussian line
of sight velocity dispersion characterized by the line width parameter
$b$~=~(2K\Tk/m~+~2$v_t^2$)$^{1/2}$, where $v_t$ is the r.m.s
(Gaussian) turbulent velocity along the line of sight. The intrinsic
line profiles were calculated assuming a minimum number of velocity
components and were then convolved with the instrumental response.
The three parameters ($N$, $v$, $b$) for each cloud were then determined by
an iterative, nonlinear, least-square fit to the observed profile.
The parameters were fine-tuned until a satisfactory fit was achieved. The main criterion of a satisfactory fit is that the r.m.s. deviation ($\sigma$) from the observed 
fit is comparable to the r.m.s. fluctuation in the stellar continuum. The uncertainties were determined by varying one parameter while keeping others unchanged until the residual between the fit and observed profile goes beyond 
3$\sigma$ of the stellar continuum.\\

For both HD~141569 and HD~157841, we started with the K~{\sc i} profile. To provide a good fit, one component was required for HD~141569 and two components for HD~157841. We started with the components determined from the K~{\sc i}
fitting and added more components to generate good fits for the other lines.
For the case of the Ca~{\sc ii} K profile, the prominent features coincide with the components seen in K~{\sc i} but we needed to include additional components for both HD~141569 and HD~157841 to generate good fits. In the case of HD~157841 for example, only one of the K~{\sc i} components coincides with the Ca~{\sc ii} K
components and we needed a minimum of 5 more components to achieve a good fit.
For the Na D$_1$ line, the central part is saturated for both stars but in the case of HD~157841 for example, the four components derived from the Ca~{\sc ii} K fitting, were sufficient to obtain a satisfactory fit. The theoretical profile was calculated using the same parameters for the Na D$_2$ line and a satisfactory fit was obtained, confirming the choice of $v$, $N$ and $b$ parameters used in the Na D$_1$ fit. 

The CH R$_{2}$ (1) line is split due to $\Lambda$-doubling in the $^2$$\Pi$ ground state of the molecule, with the two components being  separated by 1.43 \kms\  (Black \& van Dishoeck, 1988). The two $\Lambda$ doublet components have equal strength which results in a symmetric absorption-line profile in the absence of additional unresolved velocity structure. The CH line is weak toward both stars and the S/N ratios of the spectra are probably not enough to clearly see this splitting. The slight asymmetry of the profiles suggests the possible presence of unresolved cloud structure. We have however used only one component to fit the CH profile for each of the stars.  

Penprase (1993) obtained CH spectra for HD~141569 and HD~157841 at a much lower resolution of $\lambda$/$\delta$$\lambda$ $\sim$ 75,000. Although the values of the velocity centroids for the CH profiles
were not published in this paper, a comparison of our UHRF spectra with Penprase's spectra shows consistency in the CH velocities.

Figures 1 and 2 show the normalized profiles plotted against LSR velocity in the various absorption lines for HD~141569 and HD~157841 respectively. The observed profiles are shown as histograms while the best-fit model absorption profiles are shown as continuous lines. The parameters resulting from the profile fitting are listed in Tables 1 and 2 for HD~141569 and HD~157841 respectively. 

\subsection {H~{\sc i} 21-cm data}

H~{\sc i} 21-cm data from the Galactic survey made with the 
Dwingeloo 25~m telescope (Hartmann 1994, Hartmann and Burton 1997) for the 
regions around 
HD~141569 and HD~157841 in the velocity range $-$100 \kms\ to +100 \kms\ are presented in Figures 3 and 4 respectively. The H~{\sc i} observations were made using a beam width of 36$^\prime$. The mean sensitivity of the data is about 0.07K and the velocity resolution is 1.03 \kms\ . Details of the observations and data reduction procedures can be obtained from the two references mentioned above.

The top portions of Figs. 3 and 4 show the positions of the HD~141569 and HD~157841 respectively, in galactic coordinates, with respect to the positions of the H~{\sc i} observations. The lower portions of Figs. 3 and 4 show H~{\sc i} line profiles at the 9 positions spaced 0.$^\circ$5
apart and centered approximately around the two stars. We have fitted Gaussian profiles to the H~{\sc i} data and estimated the velocities and column densities for the individual line components for the regions around HD~141569 and HD~157841 which are listed in Tables 3 and 4 respectively. Column densities have been estimated using equation 8-57 from Mihalas and Binney (1981), assuming that the material is optically thin and has a spin 
temperature of 100 K. As can be seen from Figs. 3
 and 4, there are large variations in line strength from profile to profile.
The UHRF absorption line profiles are compared to the H~{\sc i} profile in Figs. 1 and 2. The top five panels in both figures, show  the 
Na D$_{1}$, Na D$_{2}$, Ca~{\sc ii} K, K~{\sc i} and CH absorption profiles respectively while H~{\sc i} profile closest to the star is shown in the lowest panel.

\section{Discussion}
\subsection{The HD~141569 sight line}
As mentioned in $\S$ 1.1, the main ISM structures in this direction are 
L~134N
and MBM~37. L~134N is a prototypical cold, dark cloud with no internal energy source and has been extensively mapped at millimetric wavelengths. The kinetic temperature in the dense region of L~134N has been derived to be \Tk\ $\sim$ 12 K by Snell (1981).  Swade (1989) has made the most detailed mapping of this cloud in the molecular emission lines of C$^{18}$O, CS, H$^{13}$CO$^{+}$, SO, NH$_3$ and
C$_3$H$_2$. The beam sizes used for these observations range from 46 to 99$^{\prime\prime}$ with a grid spacing of $\sim$ 1$^{\prime}$ and velocity resolution of 0.16 \kms\ . The emission lines in all the molecular spectra
are relatively narrow (FWHM $\sim$ 0.5 to 2 \kms\ ) and are centered around \vlsr\ ranging between 2.16 to 2.65 \kms\ with a typical uncertainty of 
 $\sim$ 0.35 \kms\ . These values are consistent with the H$_2$CO velocities
of $\sim$ 2.4$\pm$ 0.7 \kms\ at 2 mm and $\sim$ 2.8$\pm$0.9 \kms\ at 2 cm obtained by
Snell (1981) and the [C~{\sc i}] and CO velocities of $\sim$ 2.3 \kms\ obtained by Stark et al.
(1996). MBM~37, the molecular core associated with L~134N has a similar velocity
of $\sim$ 2.3 \kms\ derived from the CO(1$\rightarrow$0) observations by
Magnani et al. (1985), implying close association with L~134N. 

The dark clouds L~134 and L~134N appear to be connected to several clouds
and have been described as resembling ``beads on a string" by Magnani et al. (1996). These authors report that projected on these clouds is an H~{\sc i} filament with significantly higher LSR velocity. We checked the 21-cm 
survey of Heiles \& Habing (1974) and there is a large-scale H~{\sc i} feature extending between $l$ $\sim$ 10 to $\sim$ 340$^{\circ}$ and $b$
$\sim$ 36 to 38$^{\circ}$ and mainly at \vlsr\ $\sim$ --3 to +3 \kms\  but extending between $\sim$ --10 to +10 \kms\ .  There is no clear evidence for any H~{\sc i} at significantly higher LSR velocity on the Heiles \& Habing plots. 
Our H~{\sc i} profiles at position 5 do show a broad component at 6.9 \kms\  but we are not certain this is the H~{\sc i} feature referred to by Magnani et al. The column density for this H~{\sc i} feature is N(H~{\sc i}) = 3.53 $\times$ 10$^{20}$ $\pm$ 1.13 $\times$ 10$^{20}$ cm$^{-2}$ and it has has no counterpart in the UHRF spectra. We conclude that this H~{\sc i} component at 6.9 \kms\ must therefore lie beyond HD~141569.

\subsubsection {The UHRF spectra}

 We base our interpretation of the components seen in the UHRF spectra, within the context of the McKee \& Ostriker (1977) three-phase model. According to this model, the neutral ISM (the cold neutral medium, CNM) is contained in cold (T $\sim$ 80 K),
dense (n $\sim$ 10 to 1000 cm$^{-3}$) clouds which contain most of the mass of the ISM but have small filling factors. These cold, neutral clouds are surrounded by both a warm neutral medium (WNM) and a warm ionized medium (WIM). All these phases are embedded in a hot (T $\sim$ 10$^5$ K), highly ionized medium.

Toward HD~141569, the UHRF spectra show two components: one at \vlsr\ = --15.1 \kms\ which is seen only Ca~{\sc ii} K absorption and another feature at
$\sim$ 0 \kms\ that shows Na~{\sc i} D$_1$, D$_2$, Ca~{\sc ii} K, K~{\sc i}
and CH absorption. These components and their associations are discussed in detail in the following sections.

Despite the ultra-high resolution of 0.32 \kms\ of our observations, the $b$ values derived for the 5
components (listed in Table 1), is greater than 1.45 \kms\ . Welty et al.
(1994) made a high resolution survey of interstellar Na~{\sc i} D$_1$
lines towards 38 stars. Their spectra had a resolution of 0.5 \kms\ 
and an average S/N $\sim$ 100. They detected a total of 276 clouds with a 
median $b$ value of 0.73 \kms\ and a median log$N$ value of 11.09 cm$^{-2}$.
Hence there is the possiblity that the  $b$ values reported in our paper are overestimates of the true $b$ values of individual clouds, caused by our inability to
resolve the component structure due to the relatively low S/N of our spectra.
However, on the other hand, our earlier UHRF observations on the HD~28497 sightline (Blades et al.
1997) have S/N  comparable to the S/N of the spectra presented here. In the 
case of HD~28497, four of the 15 components detected in Na~{\sc i} had $b$ values ranging between 0.31 to 0.40 \kms\ while we do not detect any component toward HD~141569 (or HD~157841) with $b$ values $<$ 1.30 \kms\ . The minimum 
detectable column density for  Na~{\sc i} in the spectra presented in this paper is $\sim$ 2 $\times$ 10$^9$ cm$^{-2}$ which is about a factor 100 less than the column densities of the typical interstellar clouds studied by Welty et al. (1994). One possiblity is that the large $b$ value components seen toward HD~141569 (and HD~157841) are a superposition of several low column density 
($<$ 2 $\times$ 10$^{9}$ cm$^{-2}$) clouds, with overlapping velocities.
These are not the typical interstellar clouds described by Welty et al.
(1994) and it is unlikely that all the components seen in our spectra
are due to such overlapping low column density clouds. For this reason, we conclude that the $b$ values we estimate in this paper probably reflect 
the true $b$ values of the individual clouds.

\subsubsection{The --15 \kms\ component}

Seen in absorption only in Ca~{\sc ii} K, this weak component is relatively broad with  $b$ = 4.5 \kms\ . No H~{\sc i} counterpart to this component is seen at  position 5. The H~{\sc i} beam (36$^\prime$) is much larger than the absorption beam (which is infinitesimally small) and it could be that the H~{\sc i} emission associated with this weak component is lost in the general background emission. The Ca~{\sc ii} column density derived for this component is 8.11 $\times$ 10$^{11}$ cm$^{-2}$. Unfortunately, interstellar Ca~{\sc ii} column densities are very poorly correlated with quantities like N(H~{\sc i}),
N(Na~{\sc i}), N(K~{\sc i}) and E(B--V) (Hobbs, 1974) and it is not a useful exercise to derive N(H~{\sc i}) using the observed N(Ca~{\sc ii})
for this component. We can put an upper limit to the N(H~{\sc i}) of this component, corresponding to the sensitivity of our H~{\sc i} data which is
$\sim$ 10$^{18}$ cm$^{-2}$. Assuming that Ca~{\sc ii} is the dominant ionization stage of this element in the gas, this implies the linear depletion $\delta$(Ca)
= [N(Ca)/N(H)]/[N(Ca)/(N(H)]$_\odot$ $>$ 0.12, implying relatively low depletion for this component. 

In the absence of turbulence,  $b$ = 4.5 \kms\ for this component, corresponds to a kinetic temperature \Tk\ $\leq$ 47,900 K and such a warm
component may be located in a region close to the star. HD~141569 has a \vlsr\ radial velocity of --20.1 \kms\  and the association of the --15.1 \kms\ component with a region close to the star is not clear.

\subsubsection {The 0 \kms\ component}

The Na~{\sc i} D$_1$, D$_2$ velocities for this component match the Ca~{\sc ii}
K, K~{\sc i} and CH component velocities to better than 0.32 \kms\ . In the absence of turbulent motions, the kinetic temperature of an absorbing component for Na~{\sc i} is given by \Tk\ = 1393 $b$$^2$,
for Ca~{\sc ii} by \Tk\ = 2422 $b$$^2$, for K~{\sc i} by \Tk\ = 2368 $b$$^2$ and
for CH by \Tk\ = 787 $b$$^2$. If all these absorption components arise from coextensive gas where turbulent motions are absent, the $b$ values for the various species should differ systematically : $b$(Ca) should be 0.75$b$(Na), $b$(K) should be 0.77$b$(Na) and $b$(CH) should be 1.33$b$(Na).  From Table 1, it is clear that this is not the case since the $b$ values for Ca~{\sc ii} and K~{\sc i} exceed 
the $b$(Na) value and the $b$(CH) value is $>$ 1.33 $\times$ $b$(Na). 
On the other hand, if turbulent motions were dominant in the region where these absorption components arise, the $b$ values for all the species should be exactly the same and this is also not what is observed. Therefore the various components must arise from different regions of a cloud, which represent different phases of the ISM. This is similar to the interpretations by
Barlow et al. (1995) for clouds toward $\zeta$ Oph and Blades et al. (1997)
for clouds toward HD~28497.

In the absence of turbulent motions, \Tk\ (Na) = 2940 K and \Tk\ (Ca) = 8570 K
and we interpret the Na~{\sc i} component to arise from gas associated with the cold neutral medium (CNM), enveloped by the warm gas delineated by the Ca component. The K~{\sc i} component has a \Tk\ = 12,310 K. It is possible that the K~{\sc i} absorption feature has velocity structure but the S/N ratio of 10 for the spectra, has led us to estimate a large $b$(K) for this weak  absorption feature.  This issue can be resolved by higher S/N UHRF spectra in K~{\sc i}. If these future higher S/N UHRF spectra, still show no velocity structure for this weak K~{\sc i} feature, it would imply that a warm K~{\sc i} component envelopes the cooler Ca~{\sc ii} K and Na~{\sc i} components. 

The kinetic temperature obtained for the CH component is \Tk\ = 3910 K, assuming no turbulent motions. The CH feature is weak and the S/N ratio of 42 for the CH spectra is probably not sufficient to resolve the two components of the CH molecule and the $b$ value that we have obtained may include a blend of both components, resulting in a large \Tk\ value. CH spectra  at higher S/N obtained with the UHRF would be required to clarify this issue of additional velocity components. If the higher S/N spectra still show no velocity structure, then the CH component toward HD~141569 must be associated with the warm, outer halo surrounding the cold molecular core of MBM~37. A similar interpretation was made for the broad $b$ = 1.8 \kms\ CH component toward $\zeta$ Oph by Crawford et al. (1994). The CO gas associated with both L~134N and MBM~37 (refer $\S$ 3.1) is at
\vlsr\ $\sim$ 2.4 \kms\  and  association
of the 0 \kms\ component with MBM~37 would imply that HD~141569 lies beyond $\sim$ 110 pc.

We are interested in determining abundances and depletions for the entire cloud
represented by the components at $\sim$ 0 \kms\ seen in the various species. Although the various absorption components arise in different regions of the cloud, it is valid to use the column density values derived from the UHRF spectra for the various species for the purposes of studying abundances of the $\sim$ 0 \kms\ cloud.  Based on velocity agreement, we identify the H~{\sc i} component at position 5 with \vlsr\ = 0.78 \kms\ to be associated with this 0 \kms\ UHRF component. 
This component has N(H~{\sc i}) = 5.03 $\times$ 10$^{20}$ $\pm$ 5.57 $\times$ 10$^{18}$ cm$^{-2}$ (Table 4). The linear depletions derived are $\delta$(Na) = [N(Na)/N(H)]/[N(Na)/(N(H)]$_\odot$ = 0.004,
$\delta$(Ca) = [N(Ca)/N(H)]/[N(Ca)/(N(H)]$_\odot$ = 1.4 $\times$ 10$^{-4}$ and $\delta$(K) = [N(K)/N(H)]/[N(K)/(N(H)]$_\odot$ = 0.0017 implying extremely high depletions  for this cloud. High depletions imply low gas phase abundances and therefore high dust phase abundances since the missing elements are believed to be incorporated into dust, particularly in the case of calcium (Barlow \& Silk, 1977, Barlow 1978). Evidence for the presence of dust in this component
from the UHRF data is consistent with the strong 100$\mu$m emission from L~134N 
(Laureijs et al. 1991). 

The H$_2$ column densities in diffuse interstellar clouds are well correlated with the CH column densities through the relation N(H$_2$) = 2.6 $\times$
10$^7$ N(CH) with an r.m.s. scatter of 9.0 $\times$ 10$^{19}$ (Somerville \& Smith, 1989). Using this relation and N(CH) derived from the UHRF data, we obtain  N(H$_2$) = 1.51 $\times$ 10$^{20}$
$\pm$ 9.0 $\times$ 10$^{19}$ cm$^{-2}$. The H$_2$ fractional abundance, given by
$f$(H$_2$) = 2N(H$_2$)/[N(H~{\sc i}) + 2N(H$_2$)], for this 0 \kms\ cloud
is therefore $\sim$ 0.4$\pm$0.2. This $f$(H$_2$) value which is typically what is observed for the diffuse component associated with HLCs (Penprase, 1993). Both the evidence for the presence of dust and molecular gas and the H$_2$ fractional abundance value for this 0 \kms\ cloud strengthen the case for its association with MBM~37 and L~134N. 
 
 \subsection {The HD~157841 sight line}

A total of 6 components are seen toward HD~157841 in Na~{\sc i}, Ca~{\sc ii} K, K~{\sc i} and CH (Fig. 2 and Table 2) in the UHRF absorption spectra. The absorption features in the various species broadly resemble each other,
though they differ in details. 2 of these 6 components are seen only in a single species. A comparison of the UHRF spectra with the H~{\sc i} profile at position 4, located closest to HD~157841 (lowest panel of Fig. 2)
shows that the --8.8 \kms\ and 1.85 \kms\ UHRF components have H~{\sc i} 
counterparts. There is an additional H~{\sc i} feature
at 14.5 \kms\ which has no corresponding absorption feature on the UHRF spectra
and this H~{\sc i} emission must be associated with gas beyond HD~157841.
In the following sections, we discuss the individual components and their nature.

\subsubsection {The --15 \kms\ component}

This weak, broad $b$ = 4.97 \kms\  component is seen only in Ca~{\sc ii} K
absorption. Assuming no turbulent motions, the kinetic temperature  derived for this component is \Tk\ = 59,900 K and this component is likely to be associated with a region close to the star. As mentioned in $\S$ 1.2, we have been unsuccessful in finding any stellar radial velocity determinations for HD~157841 and therefore we cannot make any further comments  on the association of this cloud with a region close to the star.
This Ca~{\sc ii} component has no associated H~{\sc i} emission which could be due to the larger H~{\sc i} beam size causing weak components with small angular extent to dissappear in the general emission. 
 
\subsubsection {The --8.8 \kms\ component}

This component is seen in Na~{\sc i} and Ca~{\sc ii} K absorption with $b$(Na) $\neq$ 0.75$b$(Ca) or $b$(Na) = $b$(Ca) indicating that they do not arise from coextensive gas. Assuming no turbulent motions, \Tk\ (Na) = 2370 K
and \Tk\ (Ca) = 6595 K. We interpret the observed warm Ca~{\sc ii} component to envelope the colder Na~{\sc i} cloud core, similar to what is observed in case of $\zeta$ Oph (Barlow et al. 1995), HD~28497 (Blades et al. 1997) and the 0 \kms\ component toward HD~141569 ($\S$ 3.1.3.). The H~{\sc i} emission associated with this component has N(H~{\sc i}) = 4.89 $\times$ 10$^{19}$ $\pm$ 5.53 $\times$ 10$^{18}$ cm$^{-2}$. This results in $\delta$(Na) = 
[N(Na)/N(H)]/[N(Na)/(N(H)]$_\odot$ = 0.0078 and $\delta$(Ca) = [N(Ca)/N(H)]/[N(Ca)/(N(H)]$_\odot$ = 0.002, indicating relatively high depletions and implying the presence of dust.

\subsubsection {The --1.95 \kms\ component}

This component is seen in Ca~{\sc ii} K and K~{\sc i} absorption. The Ca~{\sc ii} feature is relatively narrow with $b$ = 1.68 \kms\ and 
assuming no turbulent motions, the kinetic temperature \Tk\ (Ca) = 6820 K. 
The K~{\sc i} feature is narrow with $b$ = 0.71 \kms\ which corresponds to a
\Tk\ (K) = 1210 K in the absence of turbulence. The warm Ca~{\sc ii} K component is interpreted as the warm envelope surrounding a cooler K~{\sc i} component.

\subsubsection {The 0.66 \kms\ / 1.85 \kms\ component}

The 1.85 \kms\ component is a strong feature seen in the UHRF spectra  in Na~{\sc i}, Ca~{\sc ii} and K~{\sc i} absorption. As noted in $\S$ 2.2,
the CH absorption line toward HD~157841 has a slight asymmetry and suggests possible unresolved cloud structure. We believe that due to this asymmetry, this 0.66 \kms\ component seen only in CH
is associated with the 1.85 \kms\ component seen in Na~{\sc i}, Ca~{\sc ii} and K~{\sc i}.

For the Na~{\sc i}, Ca~{\sc ii} and K~{\sc i} absorptions, $b$(Na) $\approx$ 0.75$b$(Ca) (see Table 2) $\approx$ 0.77$b$(K). The systematic difference in the linewidths in these species implies they arise from coextensive material where turbulent motions are not important. The kinetic temperatures corresponding to the derived $b$-values of the 1.85 \kms\ component are 5630~K for 
Na~{\sc i}, 5230~K for Ca~{\sc ii} and 6370~K for K~{\sc i}, in good agreement with each other. This component therefore appears to correspond quite well to the warm neutral medium (WNM) phase posited by the McKee and Ostriker model
of the ISM. The CH absorption 
has $b$(CH) = 3.19 \kms\ which is $>$ 1.33$b$(Na) (refer $\S$ 3.1.3.). However, as in the case of the CH spectra of HD~141569, the low S/N of the spectra may have led us to overestimate the $b$-value for this component.  

 The velocity of this  component closely matches the CO
LSR velocity of MBM~151, which is $\sim$ 0.8 \kms\ . This points to an association between this component and MBM~151 and places HD~157841 beyond MBM~151 which is at $\sim$ 200 pc (see $\S$ 1.2). In the absence of turbulence, the kinetic temperature \Tk\ (CH) derived for this component is $\sim$ 8000 K and  this molecular gas may be a warm envelope surrounding the cold core of MBM~151.

There is a strong H~{\sc i} 21-cm emission at 2.49 \kms\ which is most likely 
associated with the UHRF feature. The small velocity difference between this H~{\sc i} feature and the Na~{\sc i}, Ca~{\sc ii} and K~{\sc i} absorptions is not significant if one takes into consideration the differences in beam sizes for the H~{\sc i} and UHRF observations.
The column density of this feature is  N(H~{\sc i}) = 1.27 $\times$ 10$^{21}$ $\pm$ 2.15 $\times$ 10$^{19}$ cm$^{-2}$. 
Using the Somerville \& Smith (1989) relation ($\S$ 3.1.3.) and the UHRF derived CH column densities, we estimate N(H$_2$) = 5.02 $\times$ 10$^{20}$ $\pm$ 9 $\times$ 10$^{19}$ cm$^{-2}$. The H$_2$ fractional abundance, $f$(H$_2$), for this cloud is 0.45$\pm$0.08 which is typically what is found in HLCs (Penprase, 1993)
and which is similar to the value determined earlier in $\S$ 3.1.3. for the 
0 \kms\ component seen toward HD~141569.

The linear depletions derived for this cloud are $\delta$(Na) = [N(Na)/N(H)]/[N(Na)/(N(H)]$_\odot$ = 0.023, $\delta$(Ca) =  [N(Ca)/N(H)]/[N(Ca)/(N(H)]$_\odot$ = 2.8 $\times$ 10$^{-4}$ and $\delta$(K) = [N(K)/N(H)]/[N(K)/(N(H)]$_\odot$ = 0.0044 for this component. This cloud 
is therefore  highly depleted and has high dust phase abundances. Both the evidence for the presence of dust and the H$_2$ fractional abundance values, make the association of the 0.66 \kms\ CH component and the 1.85 \kms\ Na~{\sc i}, Ca~{\sc ii} and K~{\sc i} components with the same cloud and with MBM~151, more secure.

\subsubsection {The 4.65 \kms\ component}

This component is seen in both Na~{\sc i} and Ca~{\sc ii} absorption. $b$(Na) $\approx$ 0.75$b$(Ca) for this component and  the two features must arise from coextensive material.

\section{Conclusions}

We have made an ultra-high resolution study of the Na~{\sc i}, Ca~{\sc ii} K, K~{\sc i} and CH interstellar absorption lines toward two stars, HD~141569 and HD~157841. These absorption spectra have been compared to 21-cm 
data obtained from the Leiden/Dwingeloo H~{\sc i} survey. Both stars probe the gas in regions close to high galactic latitude molecular clouds : the HD~141569 sightline intercepts a region close to MBM~37 and L~134N while the HD~157841 sightline probes a region close to MBM~151.

The results of our investigation are as follows.

\noindent 1) Toward HD~141569, two components are seen in our UHRF spectra : one at --15 \kms\ and another at 0 \kms\ . 
The --15 \kms\ component, seen only in Ca~{\sc ii} K, is weak, broad and shows low linear depletion, $\delta$(Ca) $>$ 0.12. In the absence of turbulence, the kinetic temperature derived is \Tk\ $\le$ 47,900~K
and the cloud represented by this component may be located in a region close to 
HD~141569, whose stellar radial LSR velocity is --20.1 \kms\ .

The $\sim$ 0 \kms\ component is seen in Na~{\sc i}, Ca~{\sc ii} K, K~{\sc i} and CH absorption. The cloud represented by these components has a linear Ca depletion of $\delta$(Ca) = 1.4 $\times$ 10$^{-4}$, implying the presence of dust, consistent with strong 100 $\mu$m emission from this region, and H$_2$ fractional abundance $f$(H$_2$) = 0.4. We conclude that this  0 \kms\ cloud is associated with MBM~37 and L~134N because of the presence of dust and molecular (CH) gas at velocities close to 
CO velocities derived for the two HLCs. The Hipparcos distance of 99$\pm$8~pc for HD~141569 is comparable to the distance of 100 -- 110~pc that has been
estimated for the MBM~37 and L~134N clouds. However, our absorption line data
imply that these clouds must lie a little closer than HD~141569.

\noindent 2) A total of 6 components are seen in our UHRF spectra in Na~{\sc i}, Ca~{\sc ii} K, K~{\sc i} and CH absorption toward HD~157841. The cloud represented by the
1.85 \kms\ absorption components has a Ca linear depletion $\delta$(Ca) =
2.8 $\times$ 10$^{-4}$ and the derived H$_2$ fractional abundance is $f$(H$_2$) = 0.45. The relative $b$-values determined for the 1.85 \kms\ Na~{\sc i}, Ca~{\sc ii} and K~{\sc i} absorption lines are found to be consistent with
thermal broadening (with no significant turbulent component), corresponding to 
a kinetic temperature of \Tk\ = 5700$\pm$500~K. To our knowledge, this is the first velocity component found whose line widths can unambiguously be attributed
to pure thermal broadening.
We conclude that the 1.85 \kms\ cloud is associated with MBM~151 because of the presence of dust and molecular (CH) gas at velocities close to 
CO velocities derived for MBM~151. HD~141569 must therefore lie beyond MBM~151, estimated to be at $\sim$ 200 pc and its Hipparcos distance of 175$\pm$$\stackrel{45}{_{30}}$~pc is consistent with this conclusion.

\acknowledgements

This research has made use of the SIMBAD database, operated by the CDS
Strasbourg, France. MSS was supported by grant GO-6723 from Space Telescope Science Institute.

\clearpage 
\begin{deluxetable}{crccrccl} 
\tablenum{1}
\tablewidth{0pt}
\tablecaption{Model Fit Parameters for HD~141569}
\tablehead{
\colhead {} & \colhead{Na~{\sc i} D$_1$, D$_2$}  & \colhead Ca~{\sc ii} K &  \colhead{K~{\sc i}}  & \colhead{CH}}   
\startdata
\vlsr\tablenotemark{a} & \nodata & --15.1\tablenotemark{c}&\nodata &\nodata \nl
$N$\tablenotemark{b} & \nodata & 2.83\tablenotemark{d} & \nodata & \nodata \nl
$b$\tablenotemark{a} & \nodata & 4.5\tablenotemark{d} & \nodata & \nodata \nl
& & & & \nl
\vlsr\tablenotemark{a} & 0.07\tablenotemark{c} &--0.11\tablenotemark{c}& 0.23\tablenotemark{c}& 0.09\nl
$N$\tablenotemark{b} & 45.4\tablenotemark{c}& 1.68\tablenotemark{c} & 1.16\tablenotemark{d} & 58.0\tablenotemark{c} \nl
$b$\tablenotemark{a}  & 1.45\tablenotemark{c} & 4.45\tablenotemark{c} & 2.28\tablenotemark{d} & 2.23\tablenotemark{c} \nl
& & & & \nl

\tablenotetext{a}{\vlsr and $b$ are in units of \kms\ }
\tablenotetext{b}{$N$ is in units of 10$^{11}$cm$^{-2}$}
\tablenotetext{c}{error: $<$ 10\%}
\tablenotetext{d}{error: 10 to 30\%}
\tablenotetext{e}{error: $>$ 30\%}
\enddata
\end{deluxetable}

\clearpage 
\begin{deluxetable}{crccrccl} 
\tablenum{2}
\tablewidth{0pt}
\tablecaption{Model Fit Parameters for HD~157841}
\tablehead{
\colhead {} & \colhead{Na~{\sc i} D$_1$, D$_2$}  & \colhead Ca~{\sc ii} K &  \colhead{K~{\sc i}}  & \colhead{CH}}   
\startdata
\vlsr\tablenotemark{a} & \nodata & --14.97\tablenotemark{d} & \nodata & \nodata \nl
$N$\tablenotemark{b} & \nodata & 1.94\tablenotemark{d} & \nodata & \nodata \nl
$b$\tablenotemark{a} & \nodata & 4.97\tablenotemark{d} & \nodata & \nodata \nl
& & & & \nl
\vlsr\tablenotemark{a}  & --8.89\tablenotemark{c} & --8.7\tablenotemark{c} & \nodata & \nodata \nl
$N$\tablenotemark{b}  & 7.81\tablenotemark{c} & 2.40\tablenotemark{c} & \nodata & \nodata \nl
$b$\tablenotemark{a}  & 1.30\tablenotemark{c} & 1.65\tablenotemark{c} & \nodata & \nodata \nl
& & & & \nl
\vlsr\tablenotemark{a}  & \nodata & --2.17\tablenotemark{c} & --1.95\tablenotemark{c} & \nodata\nl
$N$\tablenotemark{b}  & \nodata & 2.09\tablenotemark{c} & 2.39\tablenotemark{c} & \nodata \nl
$b$\tablenotemark{a}  & \nodata & 1.68\tablenotemark{c} & 0.71\tablenotemark{c} & \nodata \nl
& & & & \nl
\vlsr\tablenotemark{a}  & \nodata  & \nodata & \nodata  & 0.66\tablenotemark{c}\nl
$N$\tablenotemark{b}  & \nodata & \nodata  & \nodata  & 193.0\tablenotemark{c}\nl
$b$\tablenotemark{a}  & \nodata  & \nodata  & \nodata & 3.19\tablenotemark{c} \nl
& & & & \nl
\vlsr\tablenotemark{a}  & $^{\ast}$1.85\tablenotemark{c,\hbox{$^\dagger$}} & 1.85\tablenotemark{c} & 1.85\tablenotemark{c} & \nodata \nl
$N$\tablenotemark{b}  & 605.0\tablenotemark{c} & 8.11\tablenotemark{c} & 7.53\tablenotemark{c} & \nodata \nl
$b$\tablenotemark{a}  & 2.01\tablenotemark{c} & 1.47\tablenotemark{c} & 1.64\tablenotemark{c}  & \nodata \nl
& & & & \nl
\vlsr\tablenotemark{a}  & $^{\ast}$4.65\tablenotemark{c} & 4.65\tablenotemark{c} & \nodata & \nodata \nl
$N$\tablenotemark{b}  & 5.78\tablenotemark{c} & 1.88\tablenotemark{c} & \nodata & \nodata \nl
$b$\tablenotemark{a}  & 1.91\tablenotemark{c} & 1.52\tablenotemark{c}  & \nodata & \nodata \nl

\tablenotetext{a}{\vlsr and $b$ are in units of \kms}
\tablenotetext{b}{$N$ is in units of 10$^{11}$cm$^{-2}$}
\tablenotetext{\ast}{component not well resolved in the spectra}
\tablenotetext{\dagger}{possible weak counterpart to the Ca~{\sc ii}
and K~{\sc i} at --1.95 \kms\ is hidden in this saturated Na~{\sc i} 
component}
\tablenotetext{c}{error: $<$ 10\%}
\tablenotetext{d}{error: 10 to 30\%}
\tablenotetext{e}{error: $>$ 30\%}
\enddata
\end{deluxetable}
\clearpage
 
\begin{deluxetable}{crccrccl} 
\tablenum{3}
\tablewidth{0pt}
\tablecaption{Model Fit Parameters to the HI line profiles for HD~141569}
\tablehead{
\colhead{Position no.}  & \colhead{$l$, $b$} &  
\colhead{velocity (km s$^{-1}$)}  & \colhead{FWHM (km s$^{-1}$)} &  
\colhead {N(HI) 10$^{19}$ cm$^{-2}$}}  
\startdata
1 & 3.$^\circ$5, 36.$^\circ$5 &$-$3.14$\pm$0.53&15.03$\pm$0.74 &1.06$\pm$0.06\nl
  &                           & 0.69$\pm$0.01  & 5.48$\pm$0.03 &9.39$\pm$0.07\nl
  &                           &13.52$\pm$0.31  &14.27$\pm$0.31 &1.14$\pm$0.02\nl

2 & 3.$^\circ$5, 37.$^\circ$0 &$-$0.26$\pm$0.96&19.58$\pm$1.21 &1.09$\pm$0.08\nl
  &                           &   0.59$\pm$0.01& 5.23$\pm$0.03 &9.62$\pm$0.04\nl
  &                           &  15.18$\pm$0.86&16.21$\pm$0.93 &0.86$\pm$0.09\nl

3 & 3.$^\circ$5, 37.$^\circ$5 &   0.46$\pm$0.01& 4.77$\pm$0.03 &9.74$\pm$0.05\nl
    &                         &   4.65$\pm$0.53&24.05$\pm$1.07 &1.42$\pm$0.03\nl
    &                         &  20.25$\pm$0.61& 9.98$\pm$2.78 &0.22$\pm$0.07\nl
    
4 &4.$^\circ$0, 36.$^\circ$5  &$-$3.08$\pm$0.47&14.09$\pm$0.80&17.30$\pm$1.40\nl
    &                         &   0.81$\pm$0.01& 5.43$\pm$0.04&54.36$\pm$0.69\nl
    &                         &  13.19$\pm$0.40&15.45$\pm$0.76&17.84$\pm$1.05\nl
    
5 &4.$^\circ$0, 37.$^\circ$0  &   0.78$\pm$0.02& 4.75$\pm$0.04&50.3$\pm$0.6\nl
    &                         &   6.92$\pm$0.39&25.63$\pm$0.80&35.3$\pm$1.13\nl
    
6 &4.$^\circ$0, 37.$^\circ$5  &   0.54$\pm$0.01& 4.99$\pm$0.02&47.98$\pm$0.24\nl
    &                         &   3.80$\pm$0.38&17.69$\pm$0.75&11.84$\pm$1.35\nl
    &                         &   9.66$\pm$0.48&31.52$\pm$0.75&23.20$\pm$1.37\nl

7 &4.5, 36.$^\circ$5  &$-$3.14$\pm$0.65&14.52$\pm$0.92&17.56$\pm$1.94\nl
    &                         &   0.92$\pm$0.01& 5.35$\pm$0.04&53.42$\pm$0.75\nl
    &                         &  12.80$\pm$0.60&16.45$\pm$0.95&17.70$\pm$1.48\nl
    
8 &4.$^\circ$5, 37.$^\circ$0  &   0.73$\pm$0.01& 5.20$\pm$0.03&51.37$\pm$0.50\nl
    &                         &   3.65$\pm$1.25&23.33$\pm$1.80&30.62$\pm$3.77\nl
    &                         &  18.13$\pm$3.44&19.84$\pm$4.58& 5.25$\pm$3.80\nl

9 &4.$^\circ$5, 37.$^\circ$5  &   0.60$\pm$0.01& 4.86$\pm$0.03&45.07$\pm$0.43\nl
    &                         &   2.88$\pm$0.44&22.19$\pm$1.11&27.24$\pm$0.88\nl
    &                         &  20.10$\pm$0.50&12.17$\pm$1.27& 4.71$\pm$0.95\nl

\enddata


\end{deluxetable}
\clearpage
\begin{deluxetable}{crccrccl} 
\tablenum{4}
\tablewidth{0pt}
\tablecaption{Model Fit Parameters to the HI line profiles for HD~157841}
\tablehead{
\colhead{Position no.}  & \colhead{$l$, $b$} &  
\colhead{velocity (km s$^{-1}$)}  & \colhead{FWHM (km s$^{-1}$)} &  
\colhead {N(HI) 10$^{19}$ cm$^{-2}$}}  
\startdata
1 &16.$^\circ$5, 15.$^\circ$5 &$-$0.42$\pm$0.12& 5.81$\pm$0.14&43.70$\pm$2.81\nl
  &                           &   5.04$\pm$0.13& 6.99$\pm$0.18&63.53$\pm$3.05\nl
  &                           &   5.89$\pm$0.19&27.84$\pm$0.56&51.45$\pm$0.98\nl

2 &16.$^\circ$5, 16.$^\circ$0 &$-$0.99$\pm$0.05& 3.30$\pm$0.15&10.87$\pm$0.98\nl
  &                           &   3.00$\pm$0.06& 8.54$\pm$0.14&81.44$\pm$2.04\nl
  &                           &   5.85$\pm$0.36&29.79$\pm$1.03&44.77$\pm$1.53\nl

3 &16.$^\circ$5, 16.$^\circ$5 &$-$9.75$\pm$0.39& 6.65$\pm$1.06& 6.32$\pm$0.78\nl
    &                         &   0.81$\pm$0.24&6.79$\pm$0.36&56.60$\pm$11.78\nl
    &                         &  5.03$\pm$0.99&9.14$\pm$0.74&40.40$\pm$10.93\nl
    &                         & 18.78$\pm$1.81&20.02$\pm$3.44&17.33$\pm$3.29\nl

4 &17.$^\circ$0, 15.$^\circ$5 &$-$8.74$\pm$0.4&4.86$\pm$0.63&4.82$\pm$0.54\nl
    &                         &   2.49$\pm$0.10&9.89$\pm$0.19&125.0$\pm$1.99\nl
    &                         &  14.51$\pm$1.28&22.92$\pm$2.20&21.7$\pm$2.54\nl

5 &17.$^\circ$0, 16.$^\circ$0 &$-$8.89$\pm$0.18&4.96$\pm$0.47&4.37$\pm$0.50\nl
    &                         &   0.91$\pm$0.10&6.59$\pm$0.15&62.11$\pm$7.18\nl
    &                         &   4.50$\pm$0.45&8.57$\pm$0.32&44.74$\pm$6.94\nl
    &                         &  9.01$\pm$0.81&36.78$\pm$1.35&32.44$\pm$1.71\nl

6 &17.$^\circ$0, 16.$^\circ$5 &$-$8.91$\pm$0.26&4.79$\pm$0.63&33.18$\pm$0.53\nl
    &                         &   1.28$\pm$0.08&6.59$\pm$0.12&69.41$\pm$4.73\nl
    &                         &   5.51$\pm$0.75&8.84$\pm$0.58&20.62$\pm$4.49\nl
    &                         &  10.42$\pm$1.00&38.42$\pm$1.67&30.98$\pm$1.87\nl

7 &17$^\circ$5, 15.$^\circ$5  &$-$9.27$\pm$0.12&3.85$\pm$0.30&2.90$\pm$0.27\nl
    &                         &   0.22$\pm$0.10&6.02$\pm$0.15&35.14$\pm$5.18\nl
    &                         &   3.66$\pm$0.20&7.88$\pm$0.16&72.82$\pm$5.24\nl
    &                         &   7.99$\pm$0.35&31.54$\pm$0.63&40.64$\pm$1.04\nl

8 &17.$^\circ$5, 16.$^\circ$0 &$-$8.92$\pm$0.13&4.61$\pm$0.33&4.09$\pm$0.34\nl
    &                         &   1.13$\pm$0.06&6.56$\pm$0.08&70.44$\pm$4.04\nl
    &                         &   4.88$\pm$0.39&8.55$\pm$0.30&30.47$\pm$3.91\nl
    &                         &  11.38$\pm$0.64&39.84$\pm$0.99&34.26$\pm$1.20\nl

9 &17.$^\circ$5, 16.$^\circ$5 &$-$0.78$\pm$0.04&2.76$\pm$0.12&10.05$\pm$0.88\nl
    &                         &   2.17$\pm$0.04&5.98$\pm$0.07&64.46$\pm$1.30\nl
    &                         &   3.61$\pm$0.18&22.61$\pm$0.68&44.10$\pm$0.81\nl
    &                         &  28.03$\pm$0.52&13.62$\pm$0.99& 6.48$\pm$0.55\nl

\enddata
       
\end{deluxetable}

 \clearpage

\figcaption 
{Normalized profiles (histograms) plotted against LSR velocity for HD~141569. The top 5 panels show UHRF absorption line data in the Na D$_1$,
Na D$_2$, Ca~{\sc ii} K, K~{\sc i} and CH lines respectively. The best-fit model absorption lines profiles are plotted as continuous thin lines. The lowest panel shows the H~{\sc i} profile at position 5 (refer Fig. 3 and $\S$ 2.3), the observation which is closest to the location of HD~141569. Note the presence of the strong absorption feature at 0 \kms\  which coincides with the H~{\sc I} peak. This component shows the presence of both molecular as well as atomic
hydrogen. For details see $\S$ 3.1.3.  \label{fig1}}

\figcaption 
{Normalized profiles (histograms) plotted against LSR velocity for HD~157841. The top 5 panels show UHRF absorption line data in the Na D$_1$,
Na D$_2$, Ca~{\sc ii} K, K~{\sc i} and CH lines respectively. The best-fit model absorption lines profiles are plotted as continuous thin lines. The lowest panel shows the H~{\sc i} profile at position 4 (refer Fig. 4 and $\S$ 2.3), the observation which is closest to the location of HD~157841. Note the presence of the strong absorption feature in the Na~{\sc i}, D$_1$, D$_2$, Ca~{\sc ii} K
and K~{\sc i} lines  at 1.85 \kms\  .\label{fig2}}

\figcaption 
{The top part of this figure shows the position of HD~141569 with respect to the positions of the beams used for the nine H~{\sc i} observations presented in $\S$ 2.3, in Galactic coordinates. The asterisk indicates the position of the star which falls just within the beam of position 5.
The lower part of the Figure shows the H~{\sc i} profiles plotted against LSR velocity, at the 9 positions taken from the Hartmann \& Burton (1995) survey.
Parameters of Gaussian fits to the profiles are listed in Table 3. 
\label{fig3}}

\figcaption 
{The top part of this figure shows the position of HD~157841 with respect to the positions of the beams used for the nine H~{\sc i} observations presented in $\S$ 2.3, in galactic coordinates. The asterisk indicates the position of the star which falls just within the beam of position 4.
The lower part of the Figure shows the H~{\sc i} profiles in the LSR velocity range --100 to +100 \kms\ , at the 9 positions taken from the Hartmann \& Burton (1995) survey.
Parameters of Gaussian fits to the profiles are listed in Table 4.\label{fig4}}

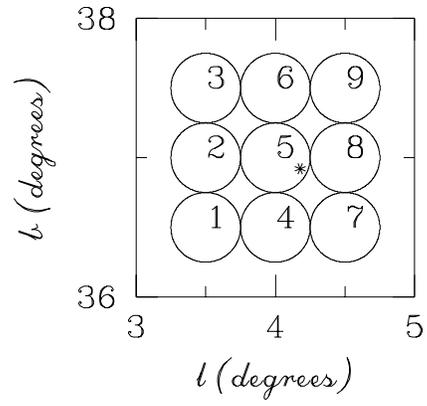
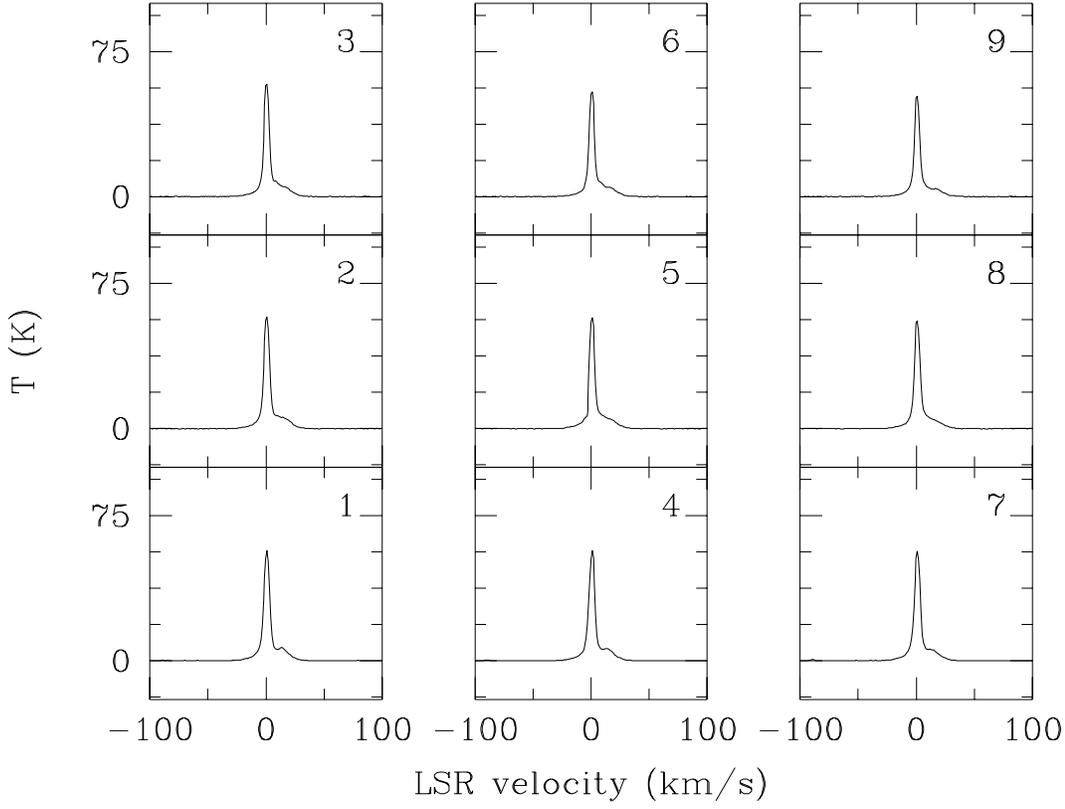
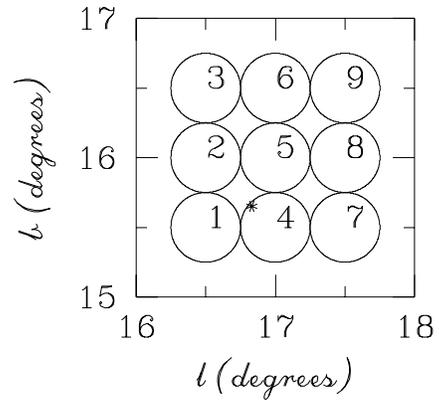
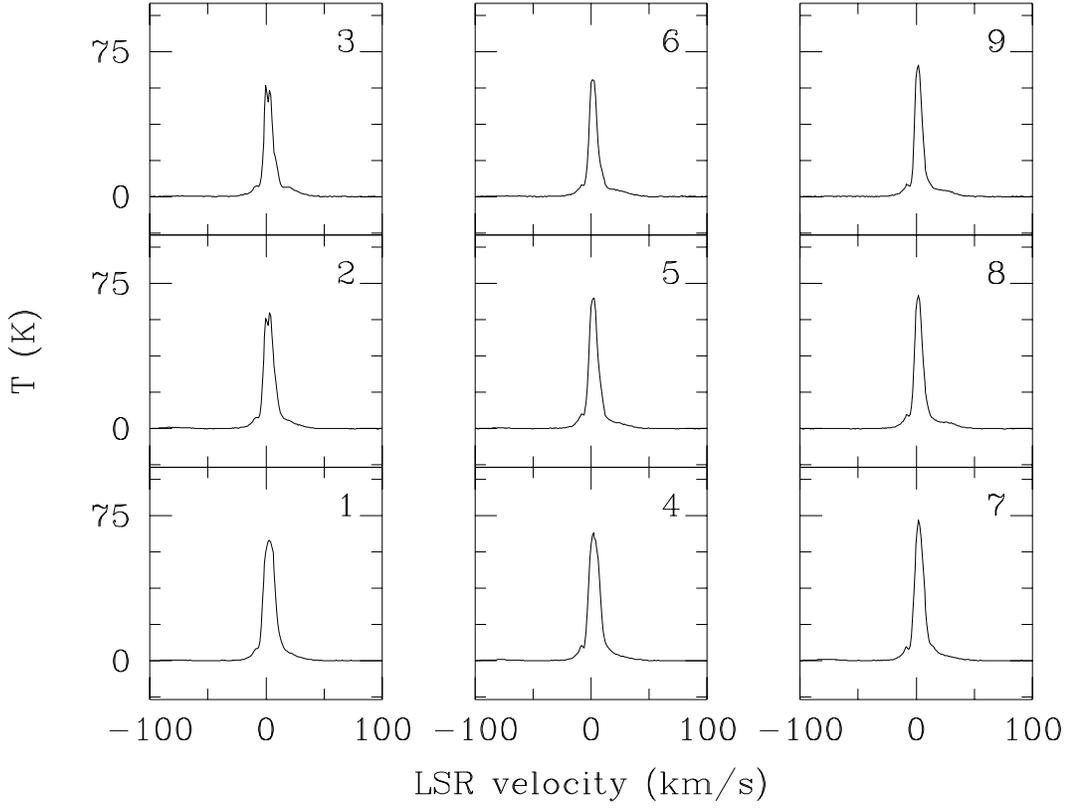
\end{document}